\documentclass[twocolumn,prl,aps]{revtex4}
\usepackage[matrix,frame,arrow]{xypic}
\usepackage{amsfonts,amsmath,amssymb}
\usepackage{graphicx,epsfig}

\input{Qcircuit}

\newcommand\beq{\begin{equation}}
\newcommand\eeq{\end{equation}}
\newcommand\bea{\begin{eqnarray}}
\newcommand\eea{\end{eqnarray}}

\newcommand{\ba}{\begin{array}}
\newcommand{\ea}{\end{array}}

\begin{document}

\author{David P. DiVincenzo$\,^1$}

\author{Panos Aliferis$\,^2$}

\title{Effective fault-tolerant quantum computation with slow measurements}

\affiliation{\vspace*{1.2ex}
             $^1$ {IBM Research Division, T.~J. Watson Research Center,
P.O. Box 218, Yorktown Heights, NY 10598}\\
             $^2$ {Institute for Quantum Information, California Institute of Technology, Pasadena, CA 91125}}

\date{\today}

\begin{abstract}

How important is fast measurement for fault-tolerant quantum
computation?  Using a combination of existing and new ideas, we
argue that measurement times as long as even 1,000 gate times or
more have a very minimal effect on
%PA2: replaced "accuracy threshold for quantum fault tolerance"
the quantum accuracy threshold. This shows that slow measurement,
which appears to be unavoidable in many implementations of quantum
computing, poses no essential obstacle to scalability.

\end{abstract}

\maketitle

Considerable progress has been made towards the physical
realization of a working quantum computer in recent years.
However, in no existing technology is there an easy pathway to
%PA: replaced "a large-scale quantum computer" with
scalability, and the main reason for this is the
stringency of the requirements for fault tolerance in quantum
computation.
%PA: added nearly
Nearly ten years ago, it was established that a quantum circuit
%PA: replaced "in which all features of its operation are noisy" with
whose components are all noisy (including storage, gate operations, state preparation and
quantum measurement) can
%PA: added efficiency
efficiently simulate
%PA: removed "the action of a noiseless quantum circuit"
any quantum computation to any desired accuracy
%PA: replaced "if"
provided the noise level is
%PA: replaced "low enough"
lower than an {\em accuracy threshold} \cite{AB:faulttol,KLZ:faulttol,Kitaev97,preskill:faulttol}. The
initial estimates of this threshold
%PA: added stochastic
for stochastic noise---around
%PA: added 10^-4
$10^{-5}$ to $10^{-4}$ error probability per
%PA: replaced "basic"
elementary operation---remain, alas, close to
the mark today.
%PA: replaced "There has been recent progress on understanding" wuth
However, recent work has investigated what special circumstances would permit the threshold value to be
higher: Notably, if qubit transport and storage are assumed to be noiseless, then
%PA: added evidence
there is evidence that the noise threshold for
%PA: added depolarizing ; removed "gate operations"
depolarizing noise exceeds $10^{-2}$\cite{knill:ftqc}.

One important
%PA: replaced "special case" with
parameter whose effect on the accuracy threshold has not been
extensively explored concerns the time it takes to complete the
%DDV3 a change here
measurement of a qubit.  Except for \cite{St03}, almost all
studies have assumed that measurements are fast---that is, that
they take no longer than
%PA: added few
%DDV: took out one or only
a few gate operation times.  This capability definitely increases
the effectiveness of error correction, since information about
errors is available promptly and any
%PA: replaced "optimal" with; why optimal?
necessary recovery operation can be applied immediately to a
logical qubit. Nevertheless, it is known that having this fast
%DDV: Moreover -> in fact
measurement capability is not necessary. In fact, measurements can
be avoided altogether and error correction can be implemented
fully coherently. The penalty paid in the stringency of the
threshold has never been quantified, but it is expected that
replacing measurement by coherent operations decreases the noise
threshold by a large amount.

In this paper we examine a scenario in which accurate quantum
measurement is possible, but is slow.  We will imagine that
measurement takes 1,000 gate operation times---a reasonable estimate
currently for spin qubits~\cite{Lieven}---but the arguments
developed here will not depend very strongly on the precise value
of this number. We find that, by combining several existing
strategies for fault-tolerant error correction with a couple of
new ``tricks",
%PA: replaced "the noise threshold values"
the accuracy threshold value is {\em barely
affected} by the speed of measurement---that is, the threshold is
hardly worse in the slow-measurement setting as compared with the
fast-measurement setting.  This diminishes one of the principal
obstacles to solid-state quantum computing, for which it is
difficult to imagine measurement times as short as gate operation
times.

%PA: added topological
Topological quantum computation~\cite{Kitaev03} aside, the best-understood route to fault-tolerant quantum computation
(FTQC) uses concatenated quantum %PA: removed "error-correcting"
codes and gate
operations applied directly to the
%PA: removed citation ~\cite{NC})
encoded data---our result is developed in this setting. We now first review some of
the principal concepts of this approach adopting language due
to~\cite{AB:faulttol}.  A quantum algorithm, laid out as a quantum
circuit, consists of a set of {\em locations}, which are
elementary components of this circuit: state preparation, one- or two-qubit gate operations (including identity ``wait'' operations when a qubit is
%PA: replaced "inactive" with
stored in memory), or qubit measurements.
Next, a ``good"
%PA: added computation
%DDV: took out error-correcting
computation quantum code is chosen.  A variety of properties make
a code ``good" for computation:  It should encode one logical
qubit in a not-very-large block of
%PA: replaced "basic" with
physical qubits
%PA: added next
while correcting some large number of errors relative to its block size.
%PA: removed "The circuit to implement error correction should be short."
A large set of
%PA: added logical
logical gate operations should be doable {\em transversally} via the application of
%PA: replaced "separate"
physical gates to each of the qubits in the code block.
%PA: is this known? proven? (It is known that there exists no code for which all necessary gates can be done transversally.)
Finally, the {\em ancilla} quantum states needed for completing
universality and for error correction should be relatively easy to
prepare in a sufficiently noiseless state.  This is typically
achieved by {\em verification}~\cite{shorver} in which the
ancilla, {\em before} being coupled to the encoded data, is
subject to tests that verify its high fidelity. These tests are
done by coupling the ancilla to other {\em verifier} ancillae,
%DDV some commas
followed by measurements on the verifier qubits, which confirm the
quality of the verified ancilla qubits or reveal the presence of
errors. In the latter case the verified ancilla is typically
rejected and the procedure starts anew.

To obtain the encoded quantum circuit, each location in the
original circuit executing the desired algorithm is replaced by a
%DDV complex -> composite
{\em rectangle}. Rectangles are composite objects consisting of a
set of locations: First, locations needed for a fault-tolerant
implementation of the ``high-level'' location (i.e., {\em logical}
state preparation, gate or measurement), followed by those
locations needed for a full error-correction cycle.  If the error
rate for elementary operations is below the accuracy threshold,
this replacement will result in an encoded circuit whose effective
noise rate is lowered
%PA: added
with respect to the original unencoded circuit. To lower the noise still further,
%PA: removed "so that an algorithm can be implemented with essentially no noise"
the replacement procedure can be repeated sufficiently many times for the locations in the encoded circuit itself. Each time, a new circuit is created which is encoded at an increasingly higher level of a {\em concatenated} quantum code.
%Even though the size of the resulting encoded circuit grows exponentially with the concatenation level, the construction is efficient: the number of concatenation levels $l$ only grows poly-logarithmically in terms of the desired accuracy since, so long as the physical noise is below the threshold, the effective error rate decreases as a double-exponential with $l$.
Although this standard concatenation procedure is
%PA: replaces "surely" ; are we sure?
not necessarily the most efficient
procedure for achieving fault tolerance,
%PA: removed "using quantum error-correcting codes" ; how else?
we will use it for the present study as its performance has
been quantified both numerically and analytically in a number of different settings.

%PA: what is the point of the next paragraph? I will try to compress it.

%The accuracy threshold should, in general, be thought of as a vector of threshold noise rates (propabilities or amplitudes depending on the postulated error model), one for each type of location in the circuit. In addition,
The standard concatenation procedure described above can be varied
and optimized in various physical settings.  For example,
%PA2: added comma after "shown that"
Knill~\cite{knill:ftqc} has shown that, in a setting where memory
and qubit transport are essentially noiseless, a very inefficient
strategy for the generation of ancilla states based on
post-selection gives a threshold around $3\times 10^{-2}$ for
depolarizing noise. On the other hand, in the more realistic
setting for contemplated solid-state implementations, where memory
has a noise level in the same range as gate operations and qubit
transport must be accomplished by noisy {\sc swap} gate
operations, a different strategy relying less on ancilla
post-selection seems to be the best. Such an approach has been
analyzed by Aliferis, Gottesman, and Preskill
(AGP)~\cite{AGP}---they find noise thresholds in this setting to
be somewhat lower than $10^{-4}$ for stochastic noise. Svore,
DiVincenzo, and Terhal (SDT)~\cite{SDT} analyze a variant of this
%DDV "to lie"
setting with qubits constrained to lie on a fixed two-dimensional
square geometry. By modifying and adapting the verification
circuits of AGP to this lattice geometry, the penalty on the
threshold found by SDT in this setting is only about a factor of
two compared with the completely unrestricted geometry of AGP.

\begin{figure}[htb]
\includegraphics[width=5.8cm,trim=0 0 0 0]{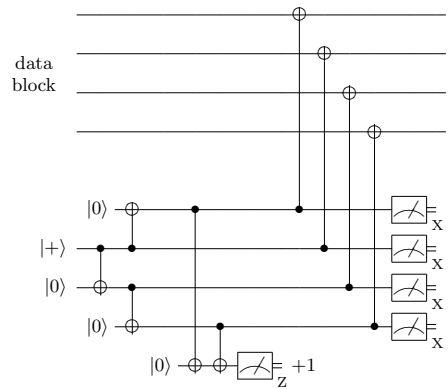}
\caption{\label{f1} Fragment of an error correction circuit in
which a ``cat state" ancilla~\protect\cite{shorver} is prepared,
verified, coupled to the data, and then measured.  The first three
controlled-NOT ({\sc cnot}) gates prepare a four-qubit
``Schr$\ddot{{\rm o}}$dinger cat"~\protect\cite{shorver} ancilla
state. The next two {\sc cnot}s and the measurement of $Z\equiv
\sigma_{\rm z}$ comprise the verification of this ancilla. In this
protocol, this measurement outcome must be known before the
verified ancilla is coupled to the data block: If the measurement
outcome is ${-}1$, the cat state is to be discarded and ancilla
preparation is to be attempted again.}
\end{figure}

%PA: rephrased next sentence
In all of this work, measurement times and gate operation times
have been assumed to be of the same order. In fact, it would seem
that the value of the accuracy threshold depends crucially on this
assumption: Most importantly, measurement is used in ancilla
verification during error correction and, the longer measurement
takes, the longer the ancilla qubits need to wait in memory while
verification is completed.
%
% DDV next two paragraphs & fig caption extensively reworked
%
The problem is illustrated by Fig.$\,$\ref{f1}, which shows a
fragment of a circuit that extracts information about errors in
the data block according to the scheme introduced by
Shor~\cite{shorver,shordiv}.
%PA2: removed the word "threshold"; I don't really see how to argue about the threshold
%PA2 which depends on what happens at higher levels of concatenation as well as the first level; this is why we say "roughly speaking"...
Roughly speaking, if measurement
takes 1,000 gate operation times, the memory %threshold
noise level
would need to be 1,000 times below the gate %threshold
noise level for the fidelity of the waiting ancilla to remain high
enough and the accuracy threshold for gate noise to stay unchanged
when slow measurement is taken in consideration.
%DDV3 a change here
%PA3: rephrased next sentence
Steane \cite{St03} has
documented such a decrease of the noise threshold with increasing
measurement time, although the effect on the threshold is not as severe
as our simple argument implies. There are some physical systems
in which the noise for qubit storage (and movement) may indeed be
very low, so that measurement-based verification can be used very
effectively to obtain high accuracy thresholds~\cite{knill:ftqc}.
%PA2: rephrased the next sentence
But in other settings (e.g., in solid-state schemes) %,
it is
expected that noise levels for gate operations, memory, and moving
will be comparable; it would seem that the threshold for FTQC
would then be severely compromised by long measurement times.

However, in this paper we show the opposite:
%PA2: added "even"; plural for settings
Even in these settings, and unlike the conclusion drawn in
\cite{St03}, the threshold is {\em hardly affected} by
%DDV3 a change here
long measurement times.  This is so because, as we discuss below (point 2.),
one can replace the non-deterministic verification protocol in Fig.$\,$\ref{f1} with a
deterministic protocol that {\em corrects} errors in the verified
ancilla.
%PA2: added next sentence
And importantly, this replacement results in an accuracy threshold comparable to that obtained with non-deterministic verification.
%
%PA2: removed the following sentence: why mention it here when we also discuss this in point 2?
%In such a deterministic protocol, the correction need not be applied directly to the
% ancilla qubits---it suffices that the necessary correction is eventually known,
% even if this knowledge becomes available only after the ancilla and the data interact.
But the full story
involves a combination of existing and new ideas, that we now
explain:

{\em 1. Use of Pauli frames.}  We did not comment above on the use
of the measurements of $X\equiv \sigma_{\rm x}$ in Fig.$\,$\ref{f1}. These
measurement bits are combined to yield the code {\em syndrome}
%PA: rephrased next: These bits and other acquired in similar EC steps are combined together to deduce the best corrective unitary operation to apply to the data qubits.
which indicates errors in the data block and the necessary recovery operation to invert them.
For all codes used in FTQC, these recovery operations are
%PA: added tensor
tensor products of single-qubit operations in the usual {\em Pauli} group.  It has been known for some time (e.g., see~\cite{knill:ftqc,AGP}) that
%PA: rephrased next: it is in many cases safe not to apply these Pauli operations immediately,
it is not necessary to directly apply these recovery operations on the data. Instead, it is sufficient to
merely record and keep track of them in a classical memory as a reference frame
defined by a Pauli rotation.  This is so because the Pauli group is closed under the action of the {\em Clifford} group:
 Pauli operators commute through gates belonging to the Clifford
%(generated by the {\sc cnot}, the Hadamard ($H$), and a few others)
% DDV: give
group to give other Pauli operators.
%PA2: rewrote next sentence "Since most of the gates in a fault-tolerant circuit" to make more precise what we imply by "most"
Since gates in a fault-tolerant circuit that determine the accuracy
threshold---most importantly, {\em all} gates needed for implementing error correction---belong to the Clifford group, the application
of the recovery operations specified by the syndrome can usually
be delayed a long time.

\begin{figure}[htb]
\includegraphics[width=6.1cm,trim=0 0 0 0]{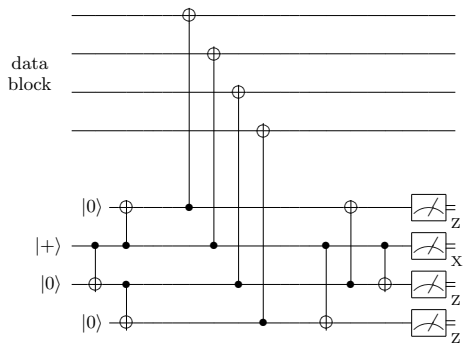}
\caption{\label{f2} The modified circuit from
Fig.$\,$\ref{f1}: ancilla verification is removed, and is
replaced by a decoding and measurement of the ancilla.}
\end{figure}

{\em 2. Ancilla decoding instead of verification.}  This is a new idea, and
requires a modification of all existing ancilla verification circuits.
%like the one in Fig.$\,$\ref{f1}.
But the modification is always simple---Fig.$\,$\ref{f2} shows the necessary change to the circuit in
Fig.$\,$\ref{f1}.
%PA: rephrased next
The reason that ancilla pre-verification before interaction with the data has previously been considered necessary is that a single fault, at certain locations in the ancilla preparation circuit, can lead to
a multi-qubit error in the ancilla state.  It has therefore always been
thought necessary to prevent such ancillae from interacting with
the data.  But, if the nature of these multi-qubit errors can
always be determined by post-processing of the ancilla after its
interaction with the data, then a suitable recovery operation
can always be devised. The decoding and
measurement of the ancilla in Fig.$\,$\ref{f2} serve to determine such a
recovery operation for the data, and this operation is again
{\em always a tensor product of single-qubit Pauli operations.}
Therefore, as in our discussion above, correction of multi-qubit errors in the ancilla can always be
delayed by incorporating the recovery operation into the Pauli frame.
The Supplementary Information gives further details of this method.

The remaining ideas are needed only to deal with these non-Clifford
operations
%PA: rephrased "Circuits implementing interesting quantum algorithms must contain them."
which, together with Clifford-group operations, complete quantum
universality. Non-Clifford operations require a different
treatment since a Pauli frame cannot be simply propagated through
them: Commuting a Pauli operator through such a gate can generally
%DDV rephrased
give an operator outside the Pauli group. For this reason, all
information determining the current Pauli frame must be known
before the application of a non-Clifford gate, so that the
restoration operation can be applied immediately before the
non-Clifford operation is implemented.

\begin{figure}[htb]
\includegraphics[width=4.2cm,trim=0 0 0 0]{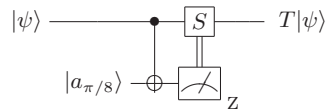}
\caption{\label{f3} Simulation of the gate $T$ using the ancilla $|a_{\pi/ 8} \rangle \equiv T \sigma_{\rm x} |0\rangle$, Clifford-group operations and measurement. The gate $S$ is performed only if the measurement outcome is ${-}1$. }
\end{figure}

We will now show that, despite this restriction, non-Clifford
gates can be executed effectively even when all measurements are
slow.
%PA: rephrased "First, we recall that in the fault-tolerant performance of non-Clifford gates, only non-Clifford state preparations (i.e., states that are not obtained from the standard $\ket{0}^{\otimes n}$ state by Clifford-group operations) are needed, not non-Clifford gate operations themselves ~\cite{shorver,preskill:faulttol,xinlan}."
First, we recall that logical non-Clifford gates are fault-tolerantly simulated using appropriate ancilla states. Non-Clifford gate operations appear in the sub-circuits preparing these ancillae, while the use of the ancillae after preparation and verification involves only Clifford-group operations
%PA: changed citation from Xinlan Zhou et al. to Gottesman and Chuang
 \cite{gate-tel}.
%PA: rephrased next
This does not immediately lead to a solution to the
measurement-time problem, as e.g. Fig.$\,$\ref{f3} illustrates.
This figure shows how to simulate the $T\equiv \exp(-i{\pi \over
8}{\sigma_{\rm z}})$ gate, with the Clifford-group gate $S = T^{\,2}$ conditioned on the
measurement outcome. Alternatively, the logical Toffoli gate could
%DDV no e.g.
be simulated, with {\sc cnot} gates being conditioned on the
measurement outcomes inside the simulation circuit
\cite{shorver,preskill:faulttol}.

%PA: rephrased the next paragraph
The simulation circuit of Fig.$\,$\ref{f3} is to be used in an
encoded form and the ancilla block will be prepared in the {\em
logical} $|a_{\pi/ 8} \rangle$ state. And, as the next step, this
circuit will also be concatenated in order to decrease the
effective noise for the logical $T$ gate to the desired level.
When the simulation of the logical $T$ gate occurs at level
$\ell$, the circuit in Fig.$\,$\ref{f3} uses a level-$\ell$
$|a_{\pi/ 8} \rangle$ ancilla. And there is a fault-tolerant
rectangle (see \cite{AGP,SDT} for the circuit) which prepares the
level-$\ell$ ancilla using level-($\ell{-}1$) $T$ gates.  In this
standard approach, these alternating replacements are iterated
until Fig.$\,$\ref{f3} is used at level $1$, where it contains
zeroth-level {\em physical} $T$ gates.

However, this circuit is clearly unusable at level $1$ if
measurements are slow: The data qubits will have to wait in memory
too long since the outcome of the measurement of level-$1$ logical
$Z$ (including all the preceding Pauli-frame information that
determines its meaning) {\em must} be known in order to decide if
the level-$1$ logical $S$ gate is to performed
%DDV added parenthetical
(this decision must be made before this qubit is involved in the
next
%PA2: added "logical"
logical {\sc cnot} in the circuit, which is usually immediately). Is
there a fix to this problem?
%PA: removed the following which is confusing...
%"If this second measurement is inappropriately applied, it will, because of the entangling action of the second CNOT, cause irreversible harm to the state of the output qubit of this circuit."

Here is the essential idea: In order to get a very low effective
error rate for the logical $T$ gate, it is only necessary that the
circuit of Fig.$\,$\ref{f3} appears at sufficiently many {\em
high} levels of concatenation. But at high levels of concatenation
there is no problem with slow measurement!
%DDV3 a change here
This is easy to see
for concatenated error correction: the gate time $t_{gate}^{(k)}$
at level $k$ of concatenation scales exponentially with $k$,
$t_{gate}^{(k)}=aC^k$ for some constants $a$ and $C$. On the other
hand, the measurement time $t_{meas}$ is
%DDV comma
the same at every level of concatenation, since logical
measurement is performed by {\em transversal} measurements on the
%DDV slight modification
physical level. So, even if $t_{meas}=1000a$ and for, e.g., $C=34$
(as in \cite{SDT}), measurement is completed in one {\rm logical}
gate time for all $k\geq k_{\rm min} = 2 \gtrsim \log_Ck$.
%DDV3 a change here
%PA3: rephrased next sentence
The more general idea is that at some level of coding, because the
effective error rate for logical Clifford-group operations
decreases quickly with coding level, the probability for a logical
error in memory in the data block in Fig.$\,$\ref{f3} can be made
sufficiently small for the total time it takes to measure the
ancilla-block qubits. This can provide a more general criterion
for determining $k_{min}$ in cases where a strategy other than
strict concatenation is used, or when $C$ is quite small ($C=5$ in
\cite{AGP}).

%DDV lower -> less
To avoid dealing with the $T$ gate at levels lower than $k_{\rm
min}$, we need an alternative to the iterative replacement
described above. One has already been suggested in the literature
(see e.g. \cite{knill:ftqc}); it is referred to as

\begin{figure}[htbp]
\includegraphics[width=7cm,trim=0 0 0 0]{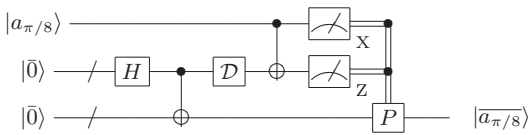}
\caption{Creation of the logical ancilla $|\overline{a_{\pi/ 8}} \rangle$ by teleportation.}
\label{f4}
\end{figure}

{\em 3. Injection by teleportation}, and this is the last ingredient
that we need.  Fig.$\,$\ref{f4} illustrates the idea:
%\marginpar{\tiny Who had this idea first? Gottesman? Steane?}
The two logical $|0\rangle$ ($|{\bar{0}}\rangle$) blocks together with the logical Hadamard and {\sc cnot} gates create a logical Bell pair  for teleportation. This logical level corresponds to $k_{\rm min}$.  Then one
of the code blocks of the Bell pair is {\em decoded} to the physical
(unencoded) level.
%PA: removed the following since it may be confusing
%One should think of this decoding as taking place one concatenation level at a time.
%Provided the noise level of the Clifford-group gates are significantly below their thresholds, the Bell pair creation and most of the decoding circuit (except for its last few steps) will be relatively noise-free.
Next, a Bell measurement is done at the physical level between a
qubit prepared in the $|a_{\pi/ 8} \rangle$ state and the decoded half of the Bell pair. As a result, up to a Pauli-frame change $P$, the output block is in the logical $|a_{\pi/ 8} \rangle$ ($|\overline{a_{\pi/ 8}} \rangle$) state as desired.
%PA:rephrased the following
The noise level on the $|\overline{a_{\pi/ 8}} \rangle$ state, which has thus been injected into the level-$k_{\rm min}$ code block, will not be much greater than that of the original $|a_{\pi/ 8} \rangle$ state and the physical noise level for Clifford operations (see e.g. \cite{SDT}).

The time required for implementing this circuit is independent of
$t_{meas}$ and, since injection occurs at level $k_{\rm min}$, all measurement outcomes will be available in one
logical gate time.  With these observations, the threshold
analyses of AGP and SDT go through essentially unchanged, so that
the accuracy threshold is not effected in the slow-measurement setting.  We say ``essentially" because
there are two changes that slow measurements make for the
threshold analysis neither of which should cause a major change
in the threshold value: i) The circuits will be changed in
detail in order to avoid ancilla verification. These changes are not
major and,
%PA: replaced "%should not very much change the location-counting analysis that leads to the threshold estimates"
because the method of ancilla decoding is more efficient than
verification, we expect the accuracy threshold to {\em improve} by
a factor of around two by this change (see the Supplementary
Information). In any case, this modification can be restricted
only to levels less than $k_{\rm min}$ since, as discussed, slow
measurement is not a problem at levels $k\geq k_{\rm min}$. ii) In
the ``local" setting of SDT, where qubits must be moved using {\sc
swap} gates on the two-dimensional lattice, extra space must be
left for qubits to be measured (since they must remain in place
for, say, 1000 time steps before they can be re-used).  This
requires an expansion of the physical patch of the lattice
occupied by one logical qubit at the lowest level of
concatenation.  We estimate that this increases the linear scale
of the computer by a small factor of around two, leading to
perhaps a factor of two decrease of the threshold. The combined
effect of (i) and (ii) leads us to expect that the accuracy
%DDV be
threshold value will be only very minimally be affected in the
slow-measurement setting.

To conclude, we have shown that
%, using a combination of existing and new ideas,
fault-tolerant quantum computation can be
implemented in such a way that, except for minor effects, slow
quantum measurements have no effect on the noise threshold at
which error correction becomes effective.  Our result does not
apply to every existing scheme for FTQC; for example, it cannot be
used in post-selected computation \cite{knill:ftqc},
%for exploiting noiseless memory and transport of qubits,
and it does not apply to the (nondeterministic) %FTQC
quantum computation scheme that is possible in a linear-optics
setting \cite{KLM}.  But in the situations generically envisioned
in many implementations of quantum computing (the solid state
ones, in particular), slow measurement will not be harmful.

We thank Daniel Gottesman and Barbara Terhal for discussions.  PA
is supported by the NSF under grant no.$\,$PHY-0456720.  DDV is
supported by the DTO through ARO contract number W911NF-04-C-0098.

\bibliographystyle{hunsrt}
%\bibliography{refs}

\cleardoublepage

\noindent {\bf \em Supplementary Information} \vskip 0.2cm

{\it The need for ancilla verification}---In the theory of quantum
fault tolerance there are two
%PA3: general-> well-known
well-known methods for performing
fault-tolerant error correction. The first, due to Shor
\cite{shorver} and DiVincenzo and Shor \cite{shordiv}, makes use
of ancillae encoded in the classical repetition code informally called ``cat'' states. The second, which is due to Steane
[A.~Steane,
%PA3: took out title
%``Active stabilisation, quantum computation and quantum state synthesis",
{\em Phys. Rev. Lett.} 78, 2252 (1997).]
and applies to the Calderbank-Shor-Steane (CSS) class of quantum
codes, uses as ancillae logical $|0\rangle$ and $|+\rangle$ states
encoded in the same code as the data. In both methods, the
preparation of these ancillae by directly executing their encoding
circuit is not in general a fault-tolerant procedure, and extra
steps are needed before the ancillae are considered pure enough to
be coupled to the data. This purification procedure, also known as
\emph{verification}, is performed by running parity checks on the
ancilla using additional ancillary verifier states.

Ancilla verification is typically done non-deterministically by simply rejecting ancillae that fail the parity checks of verification and starting anew. This procedure however is disadvantageous when measurement is slow since the verified ancilla qubits will have to be stored for long times in memory before measurements that are part of verification finish. Alternatively, verification can be done deterministically by correcting errors in the verified ancilla and thus avoiding post-selection.
%PA2: added next comment
(Note that the deterministic protocol in Fig.$\,$4 in \cite{SDT}
that does not correct errors in the ancillae is unusable when
measurement is slow: In this protocol, an element of
non-determinism exists in {\em which} ancillae are to be coupled
to the data depending on the verification measurement outcomes.)
The price for a deterministic verification procedure is a penalty in efficiency: A larger number of verifier
qubits and verification operations are needed compared to the
non-deterministic procedures (e.g., compare the deterministic
protocol in Fig.$\,$9 in \cite{preskill:faulttol} with the
non-deterministic one in Fig.$\,$11 in \cite{AGP}). And this
increase in qubit and operation overhead translates into a decrease in the accuracy threshold as compared to
non-deterministic verification---this is the reason why
non-deterministic verification has been used hitherto in almost
all threshold studies.

We will here describe how ancilla verification can be avoided
altogether and replaced by suitable decoding of the encoded
ancillae {\em after} these interact with the data during error
correction. We find that not only does this method improve the
efficiency of present schemes that use verification, but it also
leads---in the examples of interest we have studied in detail---to
an improvement in the accuracy threshold as compared even to
non-deterministic verification procedures.

%PA3: Added this new paragraph by combining pieces scattered later in the discussion
We will first discuss the case of distance-3 quantum codes
correcting arbitrary errors on any one qubit in the code block. In
the end we will discuss the generalization to codes of higher
distance. Throughout, we use the shorthand notation $X\equiv
\sigma_{\rm x}$ and $Z\equiv \sigma_{\rm z}$ and we also use
superscripts to denote the qubit on which an operator acts (e.g.,
$Z^{(i)}$ denotes a Pauli $\sigma_{\rm z}$ operator acting on the
$i$-th qubit).

{\it Ancilla decoding instead of verification}---The intuition
leading to this method is guided by a few basic observations on
how fault-tolerant error correction is implemented.

The first observation is that ancilla verification checks
particular error patterns and does not need to protect against
arbitrary multi-qubit errors. In particular, we are concerned with
faults that appear in first order in the encoding circuit and
propagate to cause multi-qubit errors at the output of the ancilla
encoder that our distance-3 code cannot correct. Verification
works by exactly trying to detect (or correct, in deterministic
schemes) these particular errors appearing in first order, with
the additional condition that the verification step should not,
also in first order, introduce multi-qubit errors in the ancilla
that is being verified.

The second observation is that the ancilla after interacting with
the data block remains, ideally, in a {\em known} logical state.
For example, after the cat state $|\bar{+}\rangle_{\rm rep} \equiv
(|00\dots 0\rangle + |11\dots 1\rangle) / \sqrt{2}$ interacts with
the data, it remains encoded in either $|\bar{+}\rangle_{\rm rep}$
or $|\bar{-}\rangle_{\rm rep}$ (the latter differs by the sign
having flipped from $+$ to $-$ in the above superposition).
Similarly in Steane's error correction method, an ancilla prepared
in the logical $|0\rangle$ state remains in the same state after
interacting with the data. This is more information than just
knowing that the ancilla state is
%PA2: rephrased the following
protected by a code---we have additional information about what the state should be.

\begin{figure}[htb]
    \begin{center}
\includegraphics[width=7.1cm,trim=0 0 0 0]{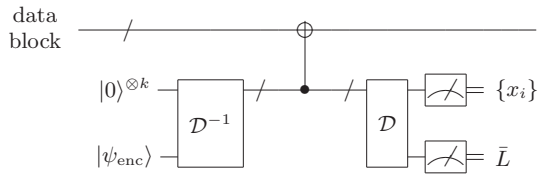}
\caption{\label{ECwoVER}
         \footnotesize{
                       Schematic of fault-tolerant error correction without ancilla verification for distance-3 codes. The ancilla is encoded by starting with ancillary qubits in a reference state (e.g., $|0\rangle^{\otimes k}$), the state to be encoded ($|\psi_{\rm enc}\rangle$), and running the unitary encoding circuit ($\mathcal{D}^{-1}$). After interacting with the data block (schematically shown as a {\sc cnot}), the ancilla is decoded using a unitary decoding circuit ($\mathcal{D}$). The final measurements yield syndrome information ($\{x_i\}$) and the eigenvalue of a logical operator ($\bar{L}$).
                       }
         }
    \end{center}
\end{figure}

The idea of the new method is to couple the encoded ancilla to the data without attempting any verification, followed by a procedure that is run afterwards on it and allows us to learn and \emph{invert} possible multi-qubit errors produced by the encoder and having propagated to the data. This procedure could be implemented in various ways, but it seems that the most efficient (and eye-pleasing) one is via a suitable decoder. A schematic of this method is given in Fig.$\,$\ref{ECwoVER}.

To prove that this error-correction (EC) method is fault tolerant,
%(for a suitable choice of the decoding circuit)
we need to show that properties $0',0-2$ in \cite{AGP} are satisfied.
Properties $0$ (EC without faults takes an arbitrary input to the
code space) and $1$ (EC without faults corrects single-qubit
errors in the data block) are true since, without faults,
verification is superfluous and the measurement outcomes after
decoding will give the necessary syndrome information to perform
ideal error correction. Properties $0'$ (EC with one fault always
produces an output deviating by at most a weight-one operator from
the code space) and $2$ (EC with one fault produces a block with
at most one error if the input has no errors) require proof. As
with the known error-correction methods that use ancilla
verification, showing that property $0'$ holds is made easier
after understanding why property $2$ is satisfied. And as we will
see, satisfying property $2$ will determine how our decoding
circuit must be designed.

What we need to show is that a single fault in the EC circuitry cannot cause more than one error in the corrected data block (which for property $2$ initially has no errors). Consider first the case where the fault occurs during the interaction with the data. Since this interaction is done with transversal gates, no gate operates on more than one qubit in the same block. Therefore a single fault cannot cause more than one error in the data. An error can also appear in the ancilla block, but this does not prevent successful correction: If cat states are used the syndrome extraction is repeated, or in Steane's method the errors in data and ancilla blocks occur always in the same qubit in the code block and a single syndrome extraction can be trusted.

The second case is when the fault occurs inside the ancilla
encoder. In general the encoding circuit has no special structure
%DDV appear in
and more than one errors can appear in the produced encoded
ancilla. The encoder can however be designed so that no logical
error appears at the output due to a single fault inside it and,
furthermore, the possible error patterns caused by all such first
order fault events are known. Of interest are multi-qubit errors which
may propagate to the data via the ancilla-data interaction.
However, letting ancilla and data interact without verification as
in Fig.$\,$\ref{ECwoVER} is permissible since the information at
the output of the following decoder will allow us to diagnose
whether such error propagation did occur and invert it. To
understand how this is possible, it is useful to think of the
decoding circuit as performing error correction while
simultaneously decoding its input block to a qubit. Since in the
case considered the decoder is fault-free, both processes are
executed ideally. The decoder then performs the mapping
\begin{equation}
\label{prop1}
E_i|\bar{0} \rangle \rightarrow |i\rangle \otimes |0\rangle \, ; \hspace{0.2cm}
E_i|\bar{1} \rangle \rightarrow |i\rangle \otimes |1\rangle \, ,
\end{equation}
\noindent where $\{E_i\}$ is the set of all single-qubit Pauli errors that our distance-3 code corrects, $|\bar{0} \rangle$ ($|\bar{1} \rangle$) is the ideal logical $|0 \rangle$ ($|1 \rangle$) state, and $i$ is syndrome information that, after decoding, indicates the error $E_i$ (if the code is not {\em perfect} and the basis $\{E_i|\bar{0} \rangle, E_i|\bar{1} \rangle \}$  does not span the whole Hilbert space,
%PA2: rephrased the following
the action of the unitary decoder can be extended to include some non-correctable errors).

Now, if we decode a $n$-qubit cat state, then measuring the qubits carrying the syndrome will allow us to learn all eigenvalues of the $Z^{(i)}Z^{(j)}$ code stabilizers for $1\leq i=j-1 \leq n-1$. Hence we can diagnose whether multi-qubit $X$ errors were produced in the encoder and can invert them by updating the Pauli frame of the data block. In Steane's method, the syndrome information will diagnose any $X$ (resp. $Z$) errors in the logical $|0\rangle$ (resp. $|+\rangle$) ancilla that propagate to the data. In addition, measuring the decoded qubit (the second tensor-product factor in Eq.$\,$(\ref{prop1}); denoted by $\bar{L}$ in Fig.$\,$\ref{ECwoVER}) will reveal the eigenvalue of the {\em logical} $Z$ (resp. $X$) operator. This resolves the ambiguity about the error causing a particular syndrome since, for $E_i$ and $E_j$ to have the same syndrome, $E_i^{\dagger} E_j = \bar{O}$ where $\bar{O}$ is either trivial (equal to the identity or $\bar{L}$) or anti-commutes with $\bar{L}$. Thus, either $E_i$ and $E_j$ are equivalent up to an element of the ancilla stabilizer, or lead to orthogonal decoded states which allows distinguishing them. Again, whenever multi-qubit errors are detected, the appropriate Pauli-frame change is done on the data block to invert them.

The final case to consider is when a single fault occurs in the
decoding circuit. In the case discussed above we were concerned
with correcting multi-qubit errors caused by a single fault in the
encoder and subsequently propagating to the data. But we also need
to guarantee that such a corrective step is not mistakenly taken
due to a single fault inside the decoder (which can cause no
errors to the data block). Thus, to satisfy property $2$, we must
ensure that no single fault inside the decoding circuit can give
the same syndrome
%PA2: added the following in parenthesis
(including the eigenvalue of the logical operator $\bar{L}$) as
any of the multi-qubit errors which a single fault inside the
encoder can produce. Constructing the decoding circuit to meet
this condition must be done with care given the chosen ancilla
encoder.

\begin{figure}[htb]
    \begin{center} \vskip 0.2cm
\includegraphics[width=8.5cm,trim=0 0 0 0]{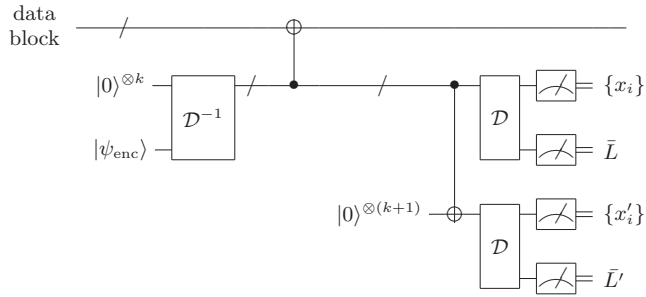}
\caption{\label{ECwoVER2}
         \footnotesize{
                       To ensure that a single fault during ancilla decoding cannot be mistaken for a fault inside the ancilla encoder leading to the same syndrome, the ancilla is encoded in the two-bit classical repetition code and then the two sub-blocks are decoded separately. If the syndrome in {\em both} sub-blocks indicates a multi-qubit $X$ error then we are confident a fault in the ancilla encoder occurred and propagated to the data. Otherwise the fault must have occurred during the decoding procedure.
                       }
         }
    \end{center}
\end{figure}

In the examples section that follows such decoder designs are
shown for some very frequently used cases: fault-tolerant
measurements using four- and seven-qubit cat states and Steane's
error-correction method for the [[7,1,3]] code. We conjecture that
an appropriate decoding circuit can be found for any distance-3
code. In any case, a less efficient but general solution is always
possible: We can further encode the ancilla after interacting with
the data in the two-bit classical repetition code and then decode
the two sub-blocks separately as shown in Fig.$\,$\ref{ECwoVER2}.
In this circuit, when the syndrome for $X$ errors at the two
%DDV agrees -> agree
sub-blocks agree, then we can be confident that, to first order,
any detected multi-qubit $X$ error has occurred during ancilla
encoding and has propagated to the data. Otherwise, if the
syndromes for $X$ errors disagree, we conclude that, again to
%DDV moved again
first order, a fault has happened in one of the two decoded
sub-blocks and no multi-qubit $X$ error has propagated to the
data. The syndrome for $Z$ errors (revealing $Z$ errors initially
in the input data block) can be obtained by taking the {\em
parity} of the syndromes at the two sub-blocks.

For such designs that satisfy property $2$,
%PA2: rephrased "it is now easy to show"
let us now discuss property $0'$.
%holds as well.
%
%PA2: took out "The important observation is"
We observe that errors that may propagate from the encoder to the data block
(e.g., $X$ errors when ancilla and data interact via a {\sc cnot}
as in Fig.$\,$\ref{ECwoVER}) are decoupled from errors that
propagate the other way around ($Z$ errors). Hence detecting and
inverting errors caused in the encoder does not interfere with the
way errors initially in the data are treated by EC. Therefore, the
%DDV reword
success of our method for dealing with single faults inside the EC
circuitry is independent of whether the data block starts in the
code space or not. This implies that EC methods which satisfy
property $0'$ when ancilla verification is used (e.g., Steane's method as discussed in \cite{AGP}), will
still satisfy it if decoding of the ancilla is performed instead as described above.

For non-perfect codes we need to worry also about replacing
verification against multi-qubit errors that do {\em not}
propagate from the ancilla to the data (e.g., $Z$ errors in
Fig.$\,$\ref{ECwoVER}). (Recall that a code is called {\em
perfect} if all possible syndromes point to correctable errors, as
is e.g. the case for the
%PA2: took out [[5,1,3]] since it has not been used in threshold calculations
Steane [[7,1,3]] and Golay [[23,1,7]] codes.)
Verification against such errors can be easily avoided by repeating the syndrome extraction: The circuit in Fig.$\,$\ref{ECwoVER} or \ref{ECwoVER2} can be repeated three times and the syndrome for $Z$ errors in the data must only be trusted if at least two of the syndromes for $Z$ errors agree.
%DDV add parenthetical
(The same applies for the $X$ error correction.)

Finally, one technical point must be addressed. With a single syndrome extraction as in
Fig.$\,$\ref{ECwoVER}, property $2$ is satisfied separately for
$X$ and $Z$ errors. That is, a single fault inside the ancilla
encoder may lead to a single $X$ and a single $Z$ error in the
output data block with the two errors acting on {\em different}
qubits inside the block. This is not a problem as long as the
subsequent logical operation does not mix $X$ and $Z$ errors, as
e.g. is the case for the logical {\sc cnot} which are transversal
for CSS codes. If however a logical $S$ gate is applied next then
we must enforce property $2$ in a stricter sense and prevent $X$
and $Z$ errors acting on different qubits at the output of EC.
This can be achieved by extracting the syndrome a second time by
running the EC circuit again. This modification will have no
effect on the accuracy threshold since the
%DDV reword
logical $S$ gate can be handled via injection by
teleportation~\cite{SDT}.

%PA2: move next one paragraph earlier
%The second technical point depends on whether the code is perfect or not. For perfect codes (i.e., codes for which any possible syndrome points to a correctable error, as is e.g. the case for the Steane [[7,1,3]] and Golay [[23,1,7]] codes) no modification is necessary. If the code is not perfect then the EC circuit in Fig.$\,$\ref{ECwoVER} or \ref{ECwoVER2} must be used three times and the syndrome for $Z$ errors in the data must only be trusted if at least two of the syndromes for $Z$ errors agree. (The same applies for the $X$ error correction.)

A similar ancilla decoding technique can replace verification when
EC uses a quantum code that corrects $t>1$ errors. Now, properties
0 -- 3 in \S 10 in \cite{AGP} are sufficient to guarantee fault
tolerance and our decoding circuit must be designed appropriately.
Most importantly, we must ensure that, with $k\leq t$ faults
inside EC, errors acting on more than $k$ qubits that may
propagate from the ancilla to the data can be diagnosed by the
subsequent ancilla decoding and inverted. This can be accomplished
by encoding the ancilla into a $t$-bit classical repetition code
before decoding each sub-block separately similar to
Fig.$\,$\ref{ECwoVER2}. For example, for the [[23,1,7]] Golay code
that corrects $t=3$ errors, ancilla decoding can replace
verification if we encode the ancilla after interaction with the
data into the $3$-bit classical repetition code [B. Reichardt,
Private communication.].

\noindent {\it Some examples}---We will now give some examples of this method in use. Let us begin with fault-tolerant syndrome measurement using four-qubit cat states, which is e.g. useful for EC with the [[5,1,3]] or [[7,1,3]] codes. The circuit for encoding, verifying and interacting these cat states with the data was shown in Fig.$\,$\ref{f1}. An alternative circuit that performs the same measurement without ancilla verification was shown in Fig.$\,$\ref{f2}.

The circuit in Fig.$\,$\ref{f2} is fault tolerant because the
error $X^{(1)}X^{(2)}$ (or its equivalent $X^{(3)}X^{(4)}$) which
may appear in the encoder in first order will be detected as it
leads to all three measurements of $Z$ after decoding giving
outcome $-1$. In addition, no single fault inside the decoding
circuit can flip all three $Z$-measurement outcomes, something
%DDV to
which would lead us to mistakenly cause a weight-two error in the
data by applying an unnecessary correction.

\begin{figure}[t]
    \begin{center}
\includegraphics[width=4.8cm,trim=0 0 0 0]{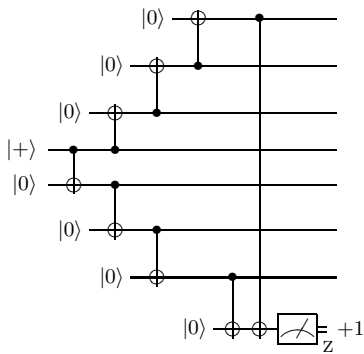}
\caption{\label{7cat}
         \footnotesize{
                       Encoding and verification of a $7$-qubit cat state. Similar to Fig.$\,$\ref{f1}, a verifier qubit is used to measure the parity $Z^{(1)}Z^{(7)}$ on the ancilla. Any of the high-weight $X$ errors ($X^{(1)}X^{(2)}$, $X^{(6)}X^{(7)}$, $X^{(1)}X^{(2)}X^{(3)}$, or $X^{(5)}X^{(6)}X^{(7)}$) created in the encoder in first order is thus detected before the ancilla interacts with the data.
                       }
         }
    \end{center}
\end{figure}

The second example is fault-tolerant measurement using seven-qubit cat states, which is e.g. needed for logical measurements that prepare ancilla needed for universality in the [[7,1,3]] code (see Figs.$\,$13 and 14 in \cite{AGP}). The circuit for encoding and verifying these cat states (Fig.$\,$14 in \cite{AGP}) is shown in Fig.$\,$\ref{7cat}.

\begin{figure}[h]
    \begin{center}
\includegraphics[width=4cm,trim=0 0 0 0]{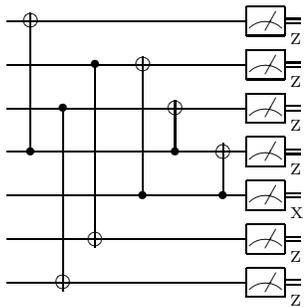}
\caption{\label{7cat_dec}
         \footnotesize{
                       The decoding circuit that replaces verification for the $7$-qubit cat state prepared as in Fig.$\,$\ref{7cat}. The outcome of the measurement of $X$ on the fifth qubit gives the eigenvalue of the measured operator on the data. As in Fig.$\,$\ref{f2}, all measurements of $Z$ give ideally outcome $+1$.
                       }
         }
    \end{center}
\end{figure}

The verification can again be omitted if, after interaction with
the data, the cat state is decoded with the circuit shown in
Fig.$\,$\ref{7cat_dec}. To understand how the outcomes of the
final measurements of $Z$ allow us to diagnose any multi-qubit $X$
errors having resulted from a single fault in the ancilla encoder, we
follow the propagation of these errors through the decoding
circuit:
\begin{equation}
\label{prop}
\begin{array}{rcl}
   X^{(1)}X^{(2)} & \rightarrow & X^{(1)}X^{(2)}X^{(6)} \\
   X^{(6)}X^{(7)} & \rightarrow & X^{(6)}X^{(7)}        \\
   X^{(1)}X^{(2)}X^{(3)} & \rightarrow & X^{(1)}X^{(2)}X^{(3)}X^{(6)}X^{(7)} \\
   X^{(5)}X^{(6)}X^{(7)} & \rightarrow & X^{(2)}X^{(4)}X^{(6)}X^{(7)} \,
\end{array}
\end{equation}
An $X$ error appearing on the right-hand side of Eq.$\,$(\ref{prop}) will result in the measurement of $Z$ on the corresponding qubit giving an outcome $-1$. We note that different initial errors propagate to different final error patterns and, hence, distinct measurement outcomes which allows distinguishing them. In addition, it is straightforward to see that no single fault inside the decoder can lead to any of the final error patterns in Eq.$\,$(\ref{prop}). So no fault inside the decoder can make us mistakenly apply a multi-qubit correction to the data.

\begin{figure}[t]
    \begin{center}
\includegraphics[width=5.2cm,trim=0 0 0 0]{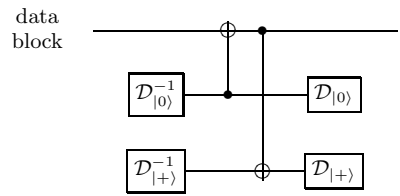}
\caption{\label{Steane_wover}
         \footnotesize{
                       Steane's EC without ancilla verification for the [[7,1,3]]. The encoding circuit for the $|0\rangle$ state ($\mathcal{D}^{-1}_{\rm |0\rangle}$) is shown separately in Fig.$\,$\ref{0enc}, and the corresponding decoder, which here includes the final measurements, ($\mathcal{D}_{\rm |0\rangle}$) in Fig.$\,$\ref{0dec}. The encoder and decoder for the $|+\rangle$ state are identical with the direction of the {\sc cnot} gates reversed and with qubit preparations and measurements performed in the conjugate bases.
                       }
         }
    \end{center}
\end{figure}

\begin{figure}[t]
    \begin{center}
\includegraphics[width=4.7cm,trim=0 0 0 0]{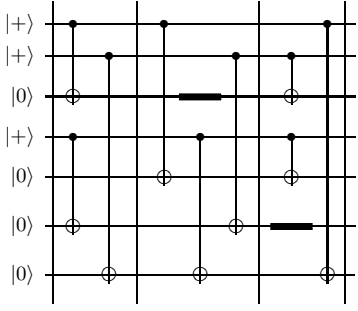}
\caption{\label{0enc}
         \footnotesize{
                       Our encoder for the logical $|0\rangle$ state in the [[7,1,3]] code. Single fault events in this circuit can lead to weight-two $X$ errors ($X^{(1)}X^{(7)}$, $X^{(2)}X^{(3)}$, or $X^{(4)}X^{(5)}$), which will propagate to the data block through the transversal {\sc cnot} gates of Fig.$\,$\ref{Steane_wover}.
                       }
         }
    \end{center}
\end{figure}

\begin{figure}[t]
    \begin{center}
\includegraphics[width=5.1cm,trim=0 0 0 0]{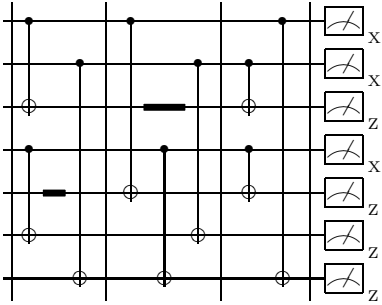}
\caption{\label{0dec}
         \footnotesize{
                       Our decoder for the logical $|0\rangle$ state in the [[7,1,3]] code corresponding to the encoder of Fig.$\,$\ref{0enc}. The measurements of $X$ yield the syndrome for $Z$ errors in the data block, while all measurements of $Z$ give ideally outcome $+1$.
                       }
         }
    \end{center}
\end{figure}

Our third example is fault-tolerant EC using Steane's method for the [[7,1,3]] code. In this method, logical $|0\rangle$ and $|+\rangle$ states are sequentially coupled to the data block to extract the syndrome information. In Steane's original scheme each ancilla is verified by either comparing two independently-encoded ancilla copies or by measuring suitable parities with extra verifier qubits (for the [[7,1,3]] code the measurement of a single parity is sufficient; see \cite{SDT}). In our variant of this method no verification is performed and after encoding the ancilla is allowed to interact with the data. The decoding circuit that is applied next to the ancilla is identical with the encoding circuit. A schematic of this EC method is shown in Fig.$\,$\ref{Steane_wover}, where the encoder ($\mathcal{D}^{-1}_{\rm |0\rangle}$) and decoder ($\mathcal{D}_{\rm |0\rangle}$) of the logical $|0\rangle$ state are shown separately in Fig.$\,$\ref{0enc} and Fig.$\,$\ref{0dec}, respectively.

To show the tolerance of this design to single faults, we first list the possible multi-qubit $X$ errors produced in first order in the encoder $\mathcal{D}^{-1}_{\rm |0\rangle}$: $X^{(1)}X^{(7)}$, $X^{(2)}X^{(3)}$, and $X^{(4)}X^{(5)}$. Propagating them through the decoding circuit of Fig.$\,$\ref{0dec} we obtain
\begin{equation}
\label{prop2}
\begin{array}{rcl}
   X^{(1)}X^{(7)} & \rightarrow & (X^{(1)})X^{(3)}X^{(5)} \\
   X^{(2)}X^{(3)} & \rightarrow & (X^{(2)})X^{(6)}X^{(7)} \hspace{0.3cm} , \\
   X^{(4)}X^{(5)} & \rightarrow & (X^{(4)})X^{(6)}X^{(7)}
\end{array}
\end{equation}
\noindent where we have put in parenthesis trivial errors acting on qubits subsequently measured in the $X$ eigenbasis. As seen in Eq.$\,$(\ref{prop2}), the first weight-two error gives a distinct syndrome from the other two, which have identical syndromes (they both flip the eigenvalues of the measurements of $Z$ on the sixth and seventh qubit). This is however to be expected, because their product $X^{(2)}X^{(3)}X^{(4)}X^{(5)}$ is in the code stabilizer and so the same recovery operator can be applied for both.

Finally, it can be easily checked that the decoder is designed so that a single fault inside it cannot lead to any of the final error patterns in Eq.$\,$(\ref{prop2}). Note that running the encoding circuit backwards would not have provided a decoding circuit with this property: Indeed, we can see that if e.g. the {\sc cnot} gate from the second to the third qubit is applied in the first time step of the decoder, then it will not be possible to distinguish whether the error $X^{(2)}X^{(3)}$ was produced by a fault in the encoder or by this decoder {\sc cnot} gate failing.

\noindent {\it Conclusion}---Avoiding ancilla verification is
helpful when measurements are slow since measurement outcomes need
not be available immediately. Furthermore, the EC circuits become
more efficient in both the number of qubits and operations compared to the EC circuits that use ancilla
verification. For example, using this new method to perform EC
inside the [[7,1,3]] {\sc cnot}-exRec in \cite{AGP} decreases the total
number of locations from $575$ to $351$ and the number of
%DDV by
ancillary qubits by half. Counting malignant pairs in the new
circuit nearly doubles the accuracy threshold lower
bound found in \cite{AGP} which changes from $2.73\times 10^{-5}$ to
$5.36\times 10^{-5}$.  This method may also prove beneficial for
fault tolerance with geometric locality constraints since the
reduction in the number of qubits will result in a smaller unit
cell and so shorter movement. Finally, we would like to comment
that it would be interesting to investigate whether a similar
ancilla decoding technique can replace verification in the
teleported error-correction method discussed in \cite{knill:ftqc}.

\end{document}